**Coexistence of unconventional spin Hall effect and antisymmetric planar Hall effect in IrO$_2$**

Yifei Yang[1,§], Sreejith Nair[2,§], Yihong Fan[1,§], Yu-Chia Chen[1], Qi Jia[1], Onri Jay Benally[1], Seungjun Lee[1], Seung Gyo Jeong[2], Zhifei Yang[2,3], Tony Low[1], Bharat Jalan[2,*], and Jian-Ping Wang[1,2,*]

1. Department of Electrical and Computer Engineering, University of Minnesota, 200 Union St. SE, Minneapolis, MN, 55455, USA

2. Department of Chemical Engineering and Materials Science, University of Minnesota, 421 Washington Ave. SE, Minneapolis, MN, 55455, USA

3. School of Physics and Astronomy, University of Minnesota, 116 Church St. SE, Minneapolis, MN, 55455, USA

§ These authors have equal contributions

* Authors to whom correspondence should be addressed: Jian-Ping Wang (jpwang@umn.edu); Bharat Jalan (bjalan@umn.edu).

**Abstract**

Crystal symmetry plays an important role in the Hall effects. Unconventional spin Hall effect (USHE), characterized by Dresselhaus and out-of-plane spins, has been observed in materials with low crystal symmetry. Recently, antisymmetric planar Hall effect (APHE) was discovered in rutile RuO$_2$ and IrO$_2$ (101) thin films, which also exhibit low crystal symmetry. In this study, we report the observation of both USHE and APHE in IrO$_2$ (111) films, using spin-torque ferromagnetic resonance (ST-FMR) and harmonic Hall measurements, respectively. Notably, the unconventional spin torque efficiency from Dresselhaus spin was more than double that of a previous report. Additionally, the temperature dependence of APHE suggests that it arises from the Lorentz force, constrained by crystal symmetry. Symmetry analysis supports the coexistence of USHE and APHE and demonstrates that both originate from the crystal symmetry of IrO$_2$ (111), paving the way for a deeper understanding of Hall effects and related physical phenomena.

Spin Hall effect is a promising approach to manipulate magnetization with high energy efficiency and speed, which is the key focus in spintronic applications.[1-3] Heavy metals and topological insulators are common spin-orbit-torque (SOT) materials due to the high spin-orbit coupling strength.[4-12] These materials, however, only possess conventional spin currents, where the spin polarization is orthogonal to both the charge current direction (X) and the spin current flow direction (Z), due to their high crystalline symmetry.[13,14] Recently, materials with low crystal or magnetic symmetry have been explored for spin currents with all three possible polarization directions: the conventional spins along Y, the Dresselhaus spins along X, and out-of-plane spins along Z, via unconventional spin Hall effect (USHE).[15-25] Specifically, the Z-spins are demonstrated to enable field-free switching of perpendicular magnets[20,22,26] and the efficiency is much better than the conventional spins.[27,28]

Planar Hall effect (PHE) is the transverse Hall response in the presence of coplanar electric and magnetic fields. In addition to ferromagnetic materials, non-magnetic materials with nontrivial Berry curvature also exhibit nonzero PHE.[29] The planar Hall signal has two components: one is even to the magnetic field (remains the same when field direction reverses), which is named as symmetric planar Hall effect (SPHE) in this work; and the other is odd to the field (changes sign when field direction reverses) and is named as antisymmetric planar Hall effect (APHE). In addition to Berry curvature, specific crystal symmetries were discovered to result in nonzero APHE in rutile $RuO_2$ and $IrO_2$ (101) films[30,31] and CuPt/CoPt heterostructure.[32] Although both USHE and APHE are closely related to the crystal symmetry, the coexistence of these effects has not yet been experimentally demonstrated.

$IrO_2$ is a Dirac nodal line semimetal, which is predicted to possess high spin Hall conductivities due to the band anticrossing in the band structure[33]. A large spin-torque efficiency was found in (001)- and (110)-oriented and polycrystalline $IrO_2$[34-36], but the spin polarization is in-plane, which is less efficient in magnetization manipulation compared to the out-of-plane spin. Recently, the X- and Z-spins were observed in (111)-oriented $IrO_2$[23] due to its lower crystal symmetry compared to the (001) and (110) orientations. Here, we observed multidirectional spin polarizations generated from the (111)-oriented single-crystal $IrO_2$ films using spin-torque ferromagnetic resonance (ST-FMR). The efficiencies from Y- and Z-spins are close to the previous report, while the efficiency of X-spin is more than twice that of the report.[23] Based on the theory of PHE in nodal line semimetals[37] and the symmetry analysis for Lorentz-induced APHE[31], we found that $IrO_2$ (111) possesses sufficiently low crystal symmetry that allows APHE. Indeed, nonzero SPHE and APHE signals are detected at various temperatures by harmonic Hall measurements. The temperature dependence of APHE suggests that the origin of APHE is Lorentz force, which agrees with the previous experimental work[31]. Symmetry analysis confirms the coexistence of USHE and APHE in $IrO_2$ (111) and suggests the potential interplay between USHE and APHE, opening the route for further understanding of the two effects and exploration of new materials with the two features.

IrO$_2$ thin films were grown on single-crystal TiO$_2$ (111) substrates using a hybrid oxide molecular beam epitaxy system. Ir was supplied in the form of a metal-organic precursor (99.9% Ir(acac)$_3$) by sublimation from a low temperature effusion cell. The Ir(acac)$_3$ source temperature was 185 °C and the growth rate was ~ 5 nm/hr. Oxygen was co-fed in the form of oxygen plasma using a radio-frequency plasma source equipped with charge deflection plates. The oxygen pressure was tuned to 5 × 10$^{-6}$ Torr and the flow rate was controlled using a mass flow controller. Prior to film growth, the substrates were cleaned sequentially in acetone, methanol and isopropanol followed by a degassing step at 200 °C in the load-lock chamber. The substrates were also annealed at the growth temperature for 20 mins in the presence of oxygen plasma prior to film growth. For IrO$_2$/CoFeB heterostructures, the CoFeB layers were deposited on IrO$_2$ by magnetron sputtering in a different vacuum chamber at room temperature, with a base pressure lower than 3 ×10$^{-8}$ Torr.

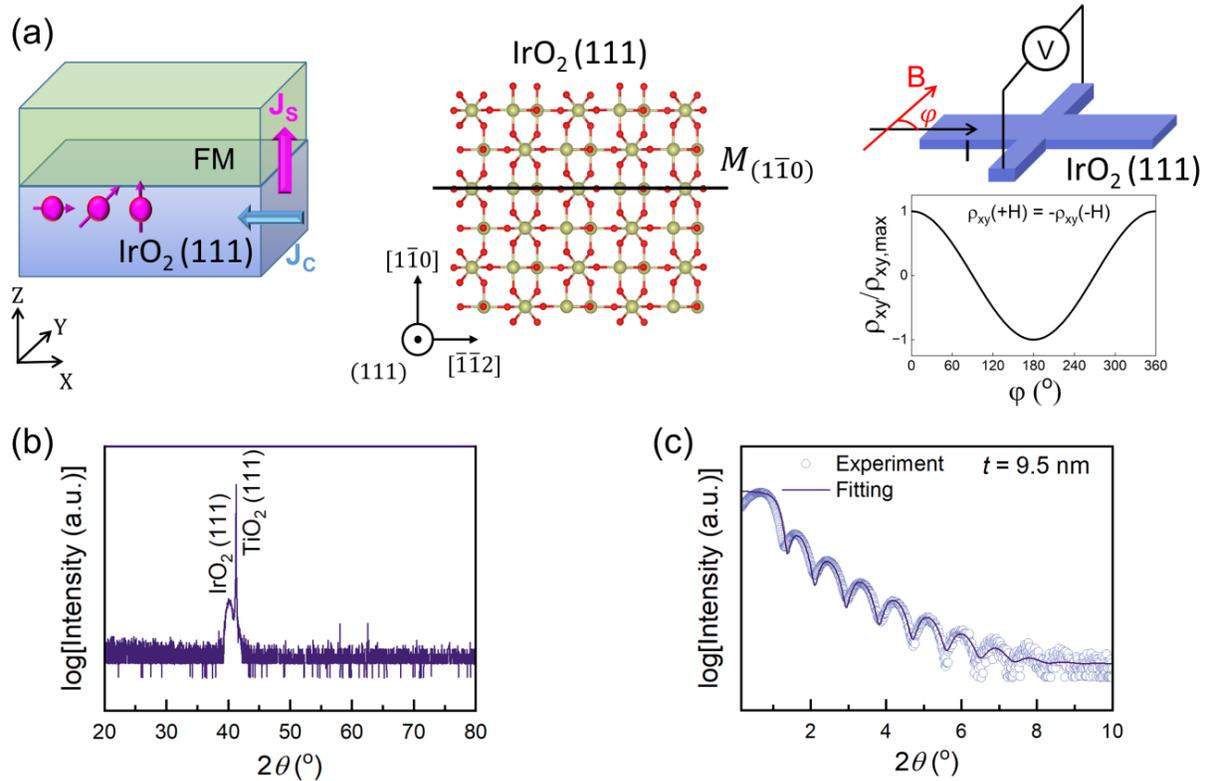

**Figure 1.** (a) Crystal structure of IrO$_2$ (111) and schematic of USHE and APHE. (b) High-resolution X-ray diffraction scan of the IrO$_2$ (111)/TiO$_2$ (111) film. (c) X-ray reflectivity measurement of the IrO$_2$ (111)/TiO$_2$ (111) sample with the fitted thickness ~ 9.5 nm.

The crystal structure of IrO$_2$ (111) is depicted in Figure 1(a), where only one mirror plane ($M_{(1-10)}$) exists. Such a low symmetry allows for the generation of both USHE and APHE, as illustrated in Fig. 1(a). Through the USHE, spin currents with spin polarizations along all X, Y, and Z directions are produced by IrO$_2$ (111). APHE can be identified by the Hall measurement, featuring a sign reversal of the Hall resistivity ($\rho_{xy}$) when the in-plane magnetic field direction is reversed. Structural characterization was performed using high-resolution X-ray diffraction using Cu K$_\alpha$ radiation. As shown in Fig. 1(b), XRD shows a phase-pure, single-crystal IrO$_2$ (111) on TiO$_2$ (111) substrate. To characterize the thickness of the IrO$_2$ film, X-ray reflectivity measurement was performed, and the result is shown in Fig. 1(c). The parameters obtained from fitting results show a film thickness of 9.5 nm and surface roughness ~ 3.6 Å, implying an extremely flat surface. Atomic force microscopy was performed on a representative IrO$_2$ (111) sample with thickness of approximately 10 nm, revealing an atomically smooth surface with root mean square roughness of 1.3 Å (Figure S1 in the supplementary material). A reference sample was also grown using similar technique on TiO$_2$ (001) substrate, and XRD again shows single crystalline IrO$_2$ (001) phase (Figure S2 in supplementary material). To verify that the deposition of CoFeB does not affect the IrO$_2$ layer, we performed XRD on the IrO$_2$ (111)/CoFeB heterostructure, as shown in Figure S3(a) in the supplementary material. No significant differences are observed compared to Fig. 1(b), confirming that the deposition of CoFeB does not damage the IrO$_2$ layer. Additionally, the rocking curve of the IrO$_2$ (111) peak in the heterostructure was measured, revealing a full width at half maximum of ~0.1° (Fig. S2(b)), which evidences the epitaxial growth of IrO$_2$ layer with high crystal quality.

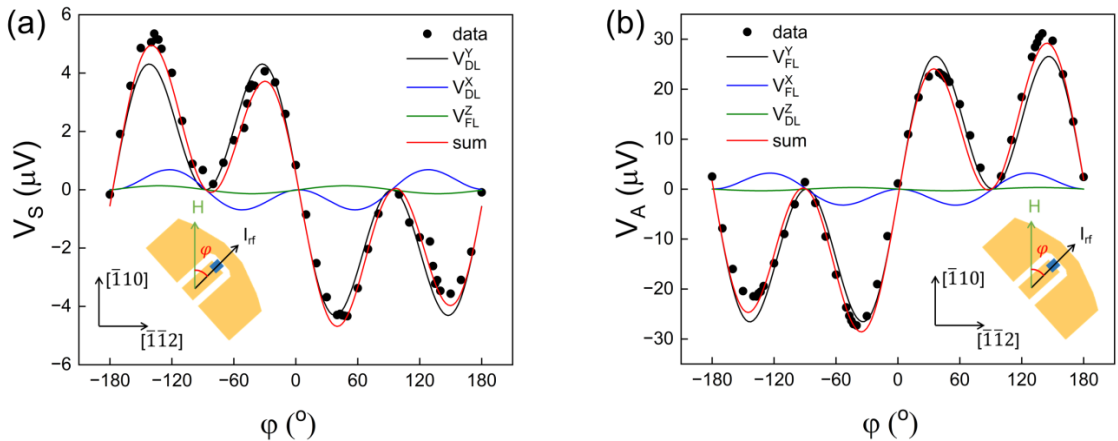

**Figure 2.** Angular dependent ST-FMR signals for IrO$_2$ (111)/CoFeB heterostructure at room temperature. The measurement schematic is shown in the inset, where the rf current is applied 45 degree from IrO$_2$ [$\bar{1}\bar{1}2$] direction. The V$_S$ (a) and V$_A$ (b) signals are fitted considering contributions from X-, Y-, and Z-spin components.

To investigate the spin Hall effect and spin polarizations generated by IrO$_2$ (111), we fabricated a heterostructure of IrO$_2$ (111)/CoFeB, with the thickness of IrO$_2$ and CoFeB of approximately 10 nm and 5 nm. ST-FMR was used to measure the spin-torque efficiencies, and the measurement schematic is shown in the inset of Figure 2 (more details in Figure S4(a) in supplementary material). A radio frequency (rf) current (I$_{rf}$) is applied to the sample and a magnetic field is swept with various orientations with respect to the I$_{rf}$, which is defined by the angle $\varphi$. The spin-induced torques cause the magnetization of CoFeB to precess and result in an oscillating anisotropic magnetoresistance. The mixing of the oscillating resistance and I$_{rf}$ produces a DC signal with resonance, which can be fitted by the sum of a symmetric (V$_S$) and an antisymmetric (V$_A$) Lorentzian.[38] An example fitting is shown in Fig. S4(b) in supplementary material.

To differentiate the contributions from spin torques by three spin directions, angular dependence measurement is carried out. The V$_S$ and V$_A$ have angular dependence:[16,19,23]

$$V_S = V_{DL}^Y \cos\varphi \sin2\varphi + V_{DL}^X \sin\varphi \sin2\varphi + V_{FL}^Z \sin2\varphi \qquad (1)$$

$$V_A = V_{FL}^Y \cos\varphi \sin2\varphi + V_{FL}^X \sin\varphi \sin2\varphi + V_{DL}^Z \sin2\varphi \qquad (2),$$

where $V_{DL}^X$, $V_{DL}^Y$, $V_{DL}^Z$ are the components from damping-like torque generated by X-, Y-, Z- spins, respectively, and $V_{FL}^X$, $V_{FL}^Y$, $V_{FL}^Z$ are their field-like counterparts. I$_{rf}$ is applied 45 degree from IrO$_2$ [$\bar{1}\bar{1}2$] direction in order to obtain non-zero X- and Z-spin simultaneously.[23] As shown in Fig. 2, V$_S$ and V$_A$ of the IrO$_2$/CoFeB heterostructure at room temperature can be well fitted by equations (1) and (2), but cannot be fitted if only considering the Y-spin. This confirms the existence of unconventional X- and Z-spins. Unconventional spin components were observed across multiple devices on the sample, as shown in Figure S5 in the supplementary material. The spin-torque efficiencies can be calculated by:[16,19]

$$\theta_{DL}^X = \frac{V_{DL}^X}{V_{FL}^Y} \frac{e\mu_0 M_S t_{FM} t_{NM}}{\hbar} \sqrt{1 + \frac{M_{eff}}{H_0}} \qquad (3)$$

$$\theta_{DL}^Y = \frac{V_{DL}^Y}{V_{FL}^Y} \frac{e\mu_0 M_S t_{FM} t_{NM}}{\hbar} \sqrt{1 + \frac{M_{eff}}{H_0}} \qquad (4)$$

$$\theta_{DL}^Z = \frac{V_{DL}^Z}{V_{FL}^Y} \frac{e\mu_0 M_S t_{FM} t_{NM}}{\hbar} \qquad (5),$$

where e is the electron charge, $\mu_0$ is vacuum magnetic permeability, $\hbar$ is reduced Planck constant, $M_S$ is saturation magnetization, $t_{FM}$ and $t_{NM}$ are the thickness of ferromagnetic CoFeB layer and nonmagnetic IrO$_2$ layer, respectively. $M_{eff}$ is the demagnetization field measured to be 1.6 Tesla obtained from Kittel formula[38] of a frequency dependence measurement (see supplementary material Fig. S2(c)). $H_0$ is the resonance field. Spin Hall conductivity is calculated by $\sigma_{DL}^{i} = \frac{\theta_{DL}^{i}}{\rho}$, where $i = X, Y, Z$, corresponding to X-, Y-, Z-spins, respectively, and $\rho$ is the charge resistivity of IrO$_2$ (111), which is 70 $\mu\Omega$ cm at room temperature. $\sigma_{DL}^{X}$, $\sigma_{DL}^{Y}$, $\sigma_{DL}^{Z}$ is estimated to be $(2.19 \pm 0.07) \times 10^4$ $\hbar/2e$ $(\Omega m)^{-1}$, $(1.38 \pm 0.04) \times 10^5$ $\hbar/2e$ $(\Omega m)^{-1}$ and $(1.51 \pm 0.21) \times 10^3$ $\hbar/2e$ $(\Omega m)^{-1}$, respectively. The $\sigma_{DL}^{Y}$ and $\sigma_{DL}^{Z}$ of the sample are similar to the values reported[23], whereas the $\sigma_{DL}^{X}$ is more than twice as large as that in the report[23], which is attributed to the outstanding crystallinity of the IrO$_2$ sample[39].

To rule out the possibility that the unconventional ST-FMR can probably originate from an out-of-plane component of the applied field[40], or asymmetric current distribution[41-44], we performed control experiments on two reference samples: Ta/CoFeB and Bi$_2$Se$_3$/CoFeB, using the identical measurement setup (Figure S6 in supplementary material). The ST-FMR signals from these samples exhibit only voltage components corresponding to the Y-spins. This unambiguously confirms that the unconventional signals observed in IrO$_2$ (111)/CoFeB are uniquely attributed to the USHE.

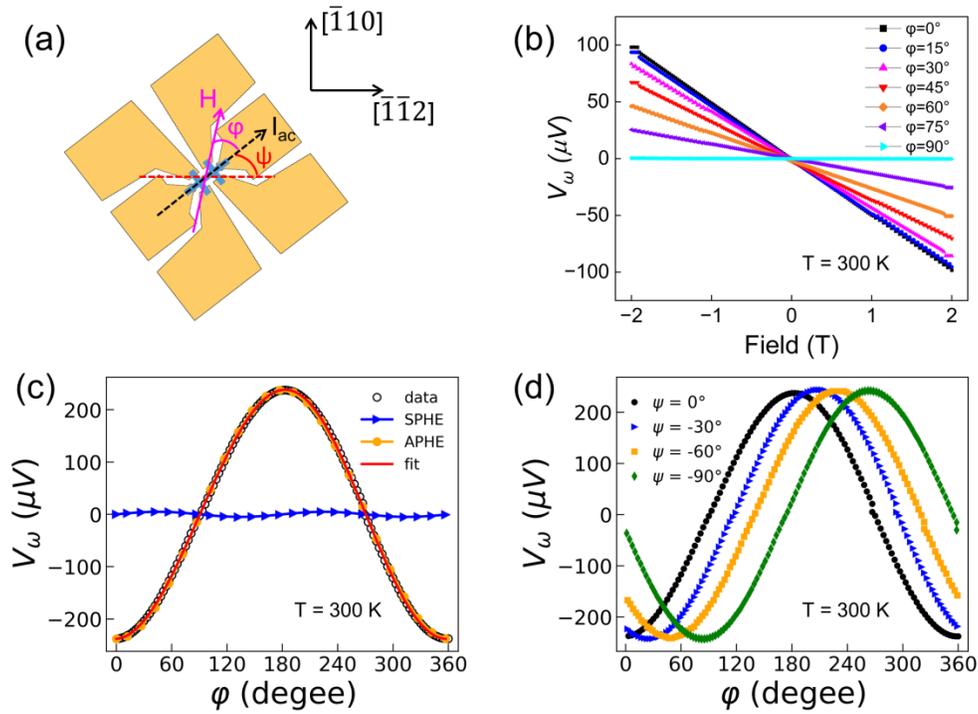

**Figure 3.** (a) Schematic of harmonic Hall measurement. (b) Hall voltage with sweeping field for different field directions of IrO$_2$ (111) sample. The current is applied along [$\bar{1}\bar{1}2$] direction. (c) Angular dependence of Hall voltage at 5 T with current along [$\bar{1}\bar{1}2$] direction, fitted as sum of SPHE and APHE. (d) Hall voltage at 5 T for four devices with different $\psi$ angles.

In addition to the USHE, we found that IrO$_2$ (111) exhibits planar Hall effect (PHE), and in particular, the APHE. To study these effects, we patterned the pure IrO$_2$ (111) with thickness of 10 nm as Hall bars, as shown in Figure 3(a). An ac current with frequency of 133.7 Hz is applied in a direction that is ψ degree away from IrO$_2$ [$\bar{1}\bar{1}2$]. A static magnetic field in the range of 500 Oe to 7 Tesla is also applied with an angle of φ with respect to the current direction in the plane of the sample. The first harmonic Hall voltage is measured at room temperature. Fig. 3(b) shows the Hall voltage with the applied current along [$\bar{1}\bar{1}2$] (ψ = 0°) for different magnetic field directions, indicating a nonzero, field-dependent PHE. Specifically, the PHE has maximum magnitude when field is along IrO$_2$ [$\bar{1}\bar{1}2$], while it disappears when field is along [$\bar{1}10$].

To better understand the contributions to the PHE, we performed the angular dependence measurements at room temperature, and the results are shown in Fig. 3(c), with the applied field to be 5 T and the current along [$\bar{1}\bar{1}2$] direction. The PHE signal is fitted to two components: $\sin 2(\varphi + \varphi_1)$ and $\cos(\varphi + \varphi_2)$, corresponding to the symmetric PHE (SPHE) and antisymmetric PHE (APHE), respectively.[29,31] SPHE is even to the magnetic field whereas APHE is odd to the magnetic field. The magnitude of APHE is much larger than SPHE and therefore APHE is the dominant contribution to the PHE signal, similar to the RuO$_2$ (101) film.[31] The Hall voltage is independent of the current frequency, as confirmed by measurements conducted at various frequencies on the same device (supplementary material Figure S7). To further explore the dependence of APHE on current direction, we measured the Hall voltage for four devices with different current directions, as shown in Fig. 3(d). The phase shift of APHE term ($\varphi_2$) is found to be identical to the current angle with respect to the IrO$_2$ [$\bar{1}\bar{1}2$] (ψ), resulting in an angular dependence of $\cos(\varphi + \psi)$ for APHE. This suggests that the APHE signal is only dependent on the magnetic field direction relative to crystal axis of IrO$_2$ (111), showing that APHE is strongly correlated to the crystal symmetry, which agrees with the other APHE phenomenon originating from crystal symmetry.[31,32] APHE and its dependence on the current directions are well reproduced in a 7 nm IrO$_2$ (111) sample as well as a IrO$_2$ (111)/CoFeB sample, as shown in Figure S8 in the supplementary material. In contrast, the SPHE term has its phase shift $\varphi_1$ as zero for all the four current directions, indicating that SPHE is only dependent on the angle between the magnetic field and current directions.

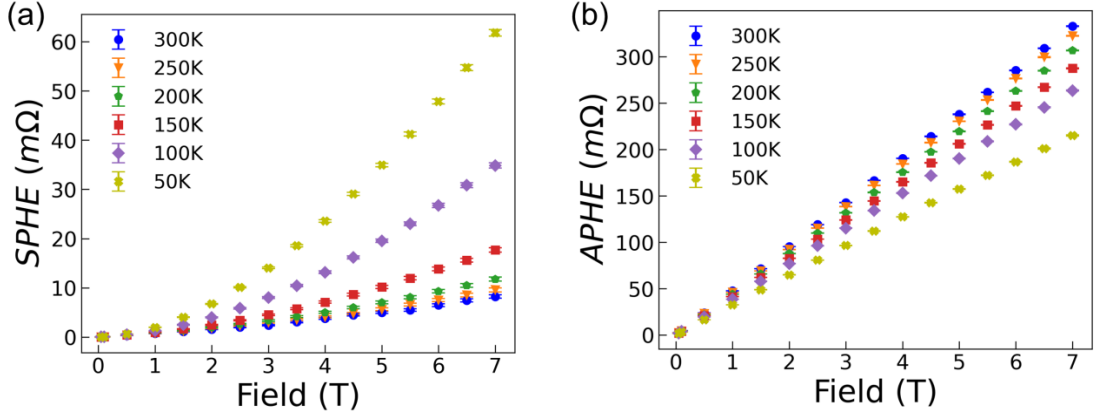

**Figure 4.** Temperature dependence of SPHE (a) and APHE (b) of $IrO_2$ (111) film, with current applied along the $[\bar{1}\bar{1}2]$ direction.

Temperature dependence study of the PHE was carried out to determine the origin of the SPHE and APHE, and the results are depicted in Figure 4. SPHE shows a quadratic dependence on the magnetic field in the small field regime but exhibits a linear dependence in the large field (Field $\geq$ 4.5 T) regime, with an example result shown in Figure S9 in supplementary material. Furthermore, it is clear that the magnitude of SPHE decreases as the temperature increases. These two characteristics have also been observed in other Dirac semimetal materials.[45-47] In contrast, APHE has a linear dependence throughout the whole field range for all temperatures. The APHE magnitudes increase slightly as the temperature increases. APHE has three possible contributions: the intrinsic orbital effect, Berry curvature, and Lorentz force. The Lorentz force contribution is independent of temperature, whereas the other two are temperature-dependent. By fitting the APHE magnitude as a function of temperature (Figure S10 in supplementary material), we found that the Lorentz force is the dominant contribution, similar to $RuO_2$[30]. The observed change in APHE magnitude may be attributed to the temperature-dependent Hall resistivity coefficients caused by variation of Fermi distribution function.[48,49] To further confirm the APHE signal originates from the low crystal symmetry of $IrO_2$ (111), we performed Hall measurements on the $IrO_2$ (001)/CoFeB and compared it to the $IrO_2$ (111)/CoFeB, as shown in Figure S11 in supplementary material. $IrO_2$ (001)/CoFeB presents a PHE-like signal that has angular dependence of $\sin2\varphi$, which originates from CoFeB, and APHE-like signal cannot be observed. In contrast, the APHE signal with an angular dependence of $\cos\varphi$ exists in $IrO_2$ (111)/CoFeB. Furthermore, no USHE can be observed in $IrO_2$ (001)/CoFeB, agreeing with previous studies.[23,34]

To understand the mechanisms of USHE and APHE controlled by symmetry, we performed the symmetry analysis on IrO$_2$. We begin our discussion with the symmetry requirements for APHE and USHE. By the definition of the Lorentz force, under an in-plane magnetic field parallel with the direction of the applied electric field ($x$), the Hall current along $y$ can be generated only if there is an electron velocity component along $z$. This scenario requires a non-zero $\sigma_{zx}$ component in the charge conductivity tensor. The conductivity tensor of $\sigma_{ij}$ is symmetrically equivalent to $v_i v_j$, where $v$ is a group velocity. Therefore, $\sigma_{zx}$ must be zero if the crystal exhibits symmetry that flips the sign of either $v_x$ or $v_z$, but not both simultaneously. These symmetries are mirror and $C_2$ rotation symmetries ($M_x, M_z, C_{2x}, C_{2z}$) and their combination with time reversal symmetry ($T$).[30,31] Materials with all these symmetries broken will exhibit APHE. Similarly, the spin Hall conductivity tensor of $\sigma_{ij}^k$ is symmetrically equivalent to $v_i s^k v_j$ where $s$ is a spin polarization, and the mirror and $C_2$ rotation symmetries also play crucial role in the shape of $\sigma_{ij}^k$.[14] Following these symmetry rules, we represented the general form of the conductivity and spin Hall conductivity tensors of (001) and (111)-orientated IrO$_2$ (space group of P4$_2$/mnm, No. 136), as shown in Table SI in supplementary material. The lattice structures of IrO$_2$ (001) and IrO$_2$ (111) are depicted in Figure S12 in supplementary material. Due to multiple mirror and $C_2$ rotation symmetries on the (001) plane, IrO$_2$ (001) does not exhibit any USHE or APHE ($\sigma_{zx}$). In contrast, the low-symmetry IrO$_2$ (111) surface has only one mirror plane ($M_{(1-10)}$), and therefore allows both USHE and APHE to exist, which agrees with the experimental observations. We note that although USHE and APHE are both correlated to the symmetry of IrO$_2$ (111), their underlying mechanisms are different. USHE is a spin-current effect that can be described by the unconventional spin Hall conductivity element, whereas APHE is a pure charge-current effect governed by Lorentz force[30,48].

In summary, we fabricated epitaxial IrO$_2$ (111) thin films by molecular beam epitaxy. USHE was observed in the IrO$_2$/CoFeB heterostructures by ST-FMR at room temperature. Notably, the SOT efficiency of X-spin was found twice that of the previous report. PHE was studied via harmonic Hall measurements at various temperatures, which is decomposed to an SPHE term and an APHE term, with the latter to be dominant. Non-zero APHE induced by Lorentz force was observed, exhibiting temperature independence and a linear dependence on the magnetic field. In contrast, SPHE displays a temperature-dependent behavior with a complex field dependence, which is commonly observed in Dirac semimetals. Symmetry analysis of IrO$_2$ (111) confirms the coexistence of the spin-related USHE and charge-related APHE, both of which are governed by the crystal symmetry. Our study paves the way for exploring material systems that exhibit both USHE and APHE and their applications in spintronics.

**Supplementary material section**

See the supplementary material for atomic force microscopy of $IrO_2$ (111), structural characterization of $IrO_2$ (001) and $IrO_2$ (111)/CoFeB, details of the ST-FMR measurement method and additional data, frequency dependence of the Hall voltage, a comparison of Hall signals between $IrO_2$ (001)/CoFeB and $IrO_2$ (111)/CoFeB, and symmetry analysis of the USHE and APHE in $IrO_2$ (111).

**Acknowledgements**

This work was supported, in part, by SMART, one of the seven centers of nCORE, sponsored by the National Institute of Standards and Technology (NIST), and by the Global Research Collaboration (GRC) Logic and Memory program, sponsored by Semiconductor Research Corporation (SRC). Films' characterization at the University of Minnesota (UMN) were supported by the Air Force Office of Scientific Research (AFOSR) through Grant No. FA9550-21-1-0025, FA9550-21-0460 and FA9550-24-1-0169. Film synthesis (S.G.J and B.J.) was supported by the U.S. Department of Energy through grant numbers DE-SC0020211, and DE-SC0024710. S.N. was supported partially by the UMN Materials Research Science and Engineering Center (MRSEC) program under Award No. DMR-2011401. Film growth was performed using instrumentation funded by AFOSR DURIP awards FA9550-18-1-0294 and FA9550-23-1-0085. Parts of this work were carried out in the Characterization Facility, University of Minnesota, which receives partial support from the NSF through the MRSEC (Award Number DMR-2011401) and the NNCI (Award Number ECCS-2025124) programs. Portions of this work were conducted in the Minnesota Nano Center, which is supported by the National Science Foundation through the NNCI under Award Number ECCS-2025124.

**Data availability**

The data that support the findings of this study are available from the corresponding author upon reasonable request.

**Coexistence of unconventional spin Hall effect and antisymmetric planar Hall effect in IrO$_2$**

Yifei Yang[1,§], Sreejith Nair[2,§], Yihong Fan[1,§], Yu-Chia Chen[1], Qi Jia[1], Onri Jay Benally[1], Seungjun Lee[1], Seung Gyo Jeong[2], Zhifei Yang[2,3], Tony Low[1], Bharat Jalan[2,*], and Jian-Ping Wang[1,2,*]

1. Department of Electrical and Computer Engineering, University of Minnesota, 200 Union St. SE, Minneapolis, MN, 55455, USA

2. Department of Chemical Engineering and Materials Science, University of Minnesota, 421 Washington Ave. SE, Minneapolis, MN, 55455, USA

3. School of Physics and Astronomy, University of Minnesota, 116 Church St. SE, Minneapolis, MN, 55455, USA

§ These authors have equal contributions

* Authors to whom correspondence should be addressed: Jian-Ping Wang (jpwang@umn.edu); Bharat Jalan (bjalan@umn.edu).

**Contents:**

1. Atomic force microscopy of IrO$_2$ (111)

2. X-ray diffraction of IrO$_2$ (001) and IrO$_2$ (111)/CoFeB

3. ST-FMR of IrO$_2$ (111)/CoFeB heterostructure and reference samples

4. Hall measurements of IrO$_2$ and IrO$_2$/CoFeB samples

5. Symmetry analysis of USHE and APHE of IrO$_2$

**References**

## 1. Atomic force microscopy of IrO$_2$ (111)

Atomic force microscopy was performed on a representative IrO$_2$ (111) sample with thickness of approximately 10 nm, with the top-view and side-view images shown in Figure S1 (a) and (b), respectively. The root mean square roughness was measured to be 1.3 Å, confirming an atomically smooth surface.

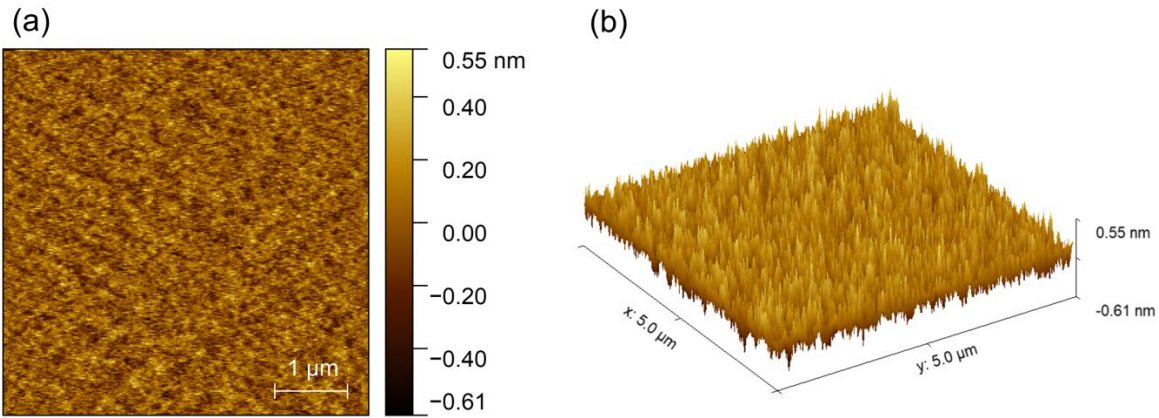

Figure S1. Atomic force microscopy images of a 10 nm IrO$_2$ (111) with root mean square roughness of 1.3 Å.

## 2. X-ray diffraction of IrO$_2$ (001) and IrO$_2$ (111)/CoFeB

High-resolution X-ray diffraction was performed on the IrO$_2$ (001) sample grown on single-crystal TiO$_2$ (001) substrate by MBE, as shown in Figure S2. Similar to IrO$_2$ (111) shown in Figure 1(a) in the main text, the IrO$_2$ (001) is also phase pure and single-crystal.

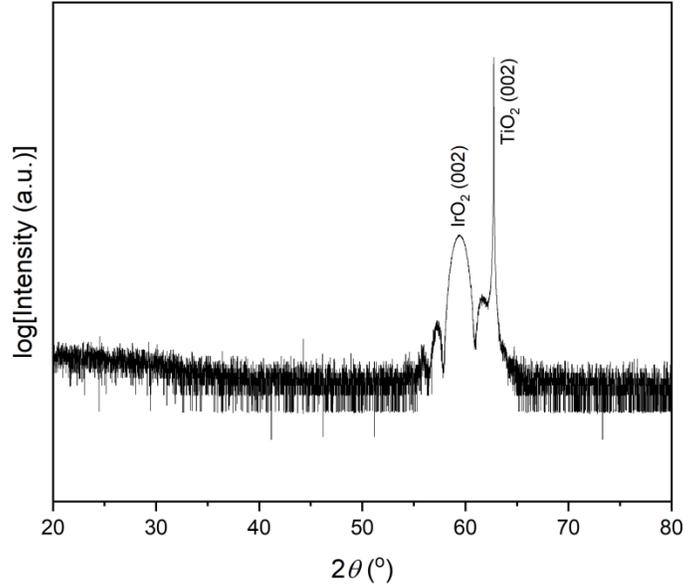

Figure S2. XRD of IrO$_2$ (001)/TiO$_2$ (001) sample.

High-resolution X-ray diffraction of IrO$_2$ (111) (~ 9.5 nm)/CoFeB (5 nm) is shown in Figure S3(a). No significant differences are observed compared to Figure 1(b), confirming that CoFeB deposition does not damage the IrO$_2$ layer. Rocking curve of the IrO$_2$ (111) peak in the heterostructure exhibits a full width at half maximum of ~0.1° (Fig. S3(b)), indicating high crystal quality.

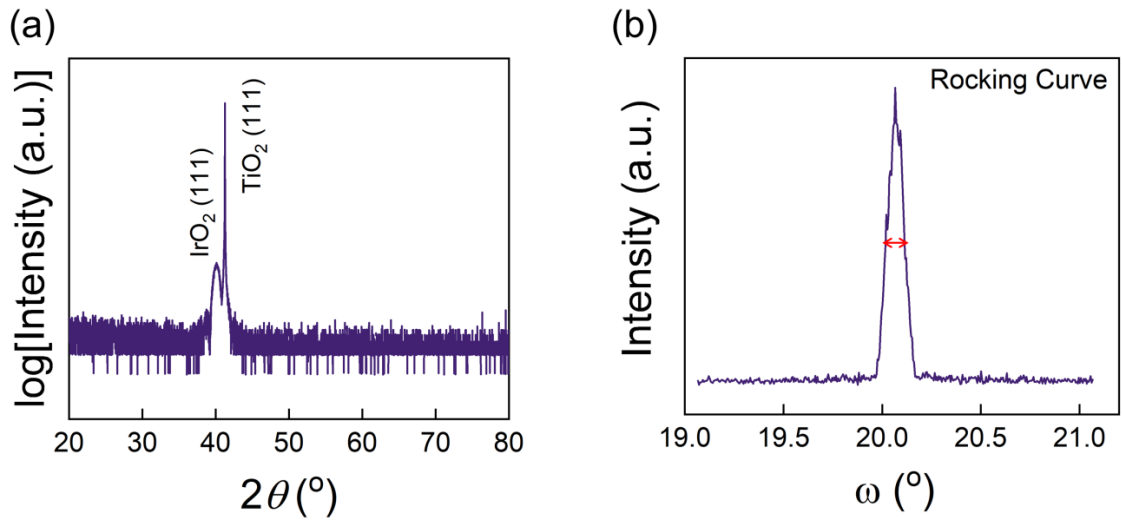

Figure S3. (a) XRD of IrO$_2$ (111) (~ 9.5 nm)/CoFeB (5 nm) heterostructure. (b) Rocking curve of the IrO$_2$ (111) peak in the heterostructure with full width at half maximum of ~0.1°.

## 3. ST-FMR of $IrO_2$ (111)/CoFeB heterostructure and reference samples

The schematic of ST-FMR measurement is shown in Figure S4(a). $IrO_2$ (111)/CoFeB heterostructure was patterned to be a strip and electrodes were deposited on top. The rf current ($I_{rf}$) was applied using a signal generator along the strip that oriented along the direction that is 45 degree angle from $IrO_2$ $[\bar{1}\bar{1}2]$. An external magnetic field is applied in the plane of the device that has an angle of φ with respect to $I_{rf}$. The DC voltage caused by the magnetization oscillation and $I_{rf}$ was measured using a nanovoltmeter. Fig. S4(b) shows an example of ST-FMR spectrum of the $IrO_2$ (111)/CoFeB heterostructure. The signal was fitted as the sum of symmetric and an antisymmetric Lorentzian, with the expression of $F_S(H_{ext}) = \Delta^2/(\Delta^2 + (H_{ext} - H_0)^2)$ and $F_A(H_{ext}) = F_S(H_{ext})(H_{ext} - H_0)/\Delta$, where $H_{ext}$ is the external magnetic field, $H_0$ is the resonance field and $\Delta$ is linewidth. Demagnetization field was calculated using the frequency-dependent resonance field data, which is shown in Fig. S4(c). The data was fitted by Kittel formula $f = \gamma/2\pi\sqrt{H_0(H_0 + 4\pi M_{eff})}$, where $f$ is resonance frequency, $\gamma$ is gyromagnetic ratio, and $4\pi M_{eff}$ is the demagnetization field.

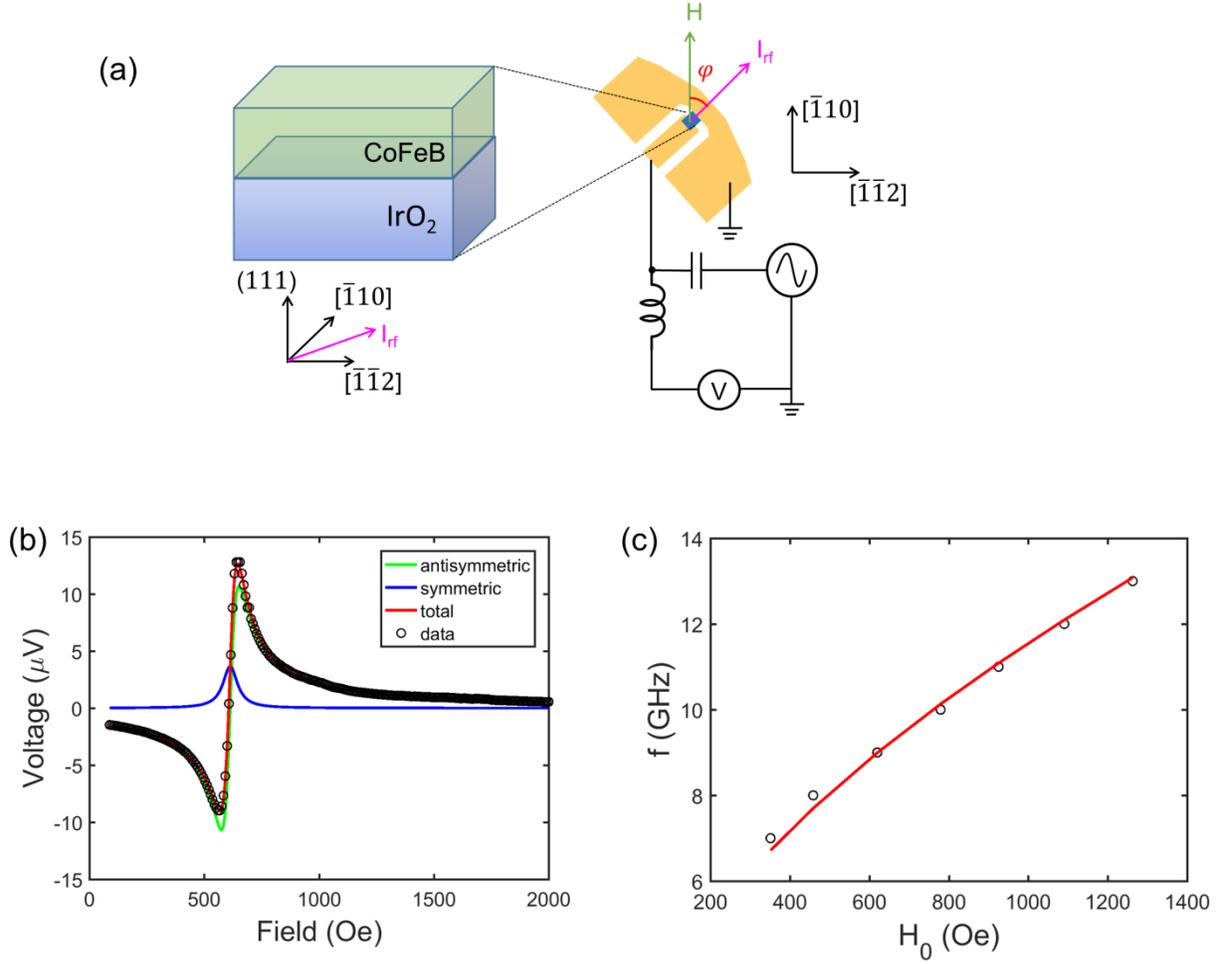

Figure S4. (a) Schematic of ST-FMR measurement. (b) ST-FMR signal of $IrO_2$ (111)/CoFeB sample at 9 GHz with external field at 45 degree from the $I_{rf}$. The signal is fitted as the sum of symmetric and antisymmetric Lorentzian. (c) Resonance frequency as a function of resonance field $H_0$, fitted by Kittel formula.

ST-FMR measurements were performed on three additional devices with the same current directions as the one used for Figure 2(b) in the $IrO_2$ (111)/CoFeB heterostructure. The resulting signals, shown in Figure S5(a), exhibit similar shapes to the one in Figure 2(b) of the main text. The calculated spin Hall conductivities show minimal variations (Fig. S5(b)), confirming the reproducibility of USHE in $IrO_2$ (111)/CoFeB.

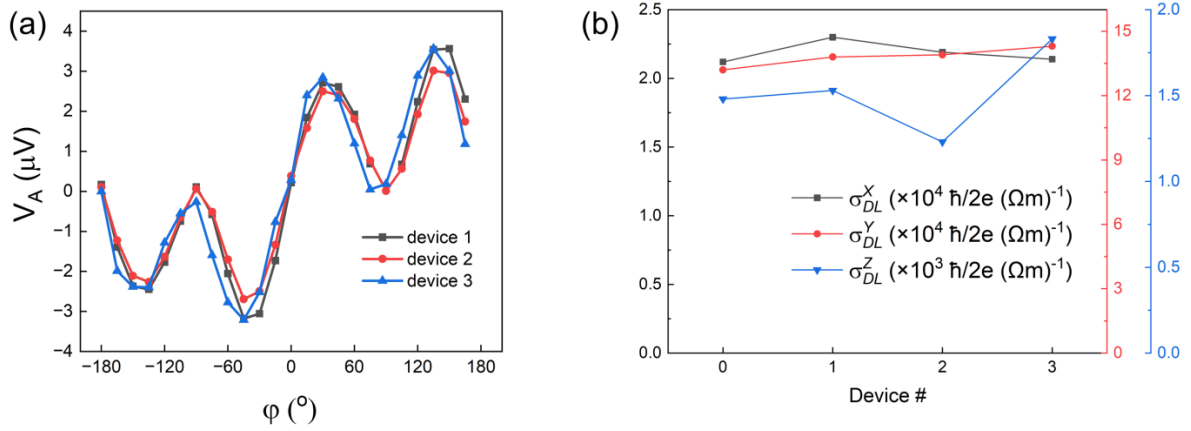

Figure S5. (a) ST-FMR signals of three different devices of the IrO$_2$ (111)/CoFeB heterostructure. (b) Calculated spin Hall conductivities of four different devices. Device #0 is the one used for Figure 2 in the main text.

ST-FMR measurements were performed on Ta (5 nm)/CoFeB (5 nm) and Bi$_2$Se$_3$ (5 nm)/CoFeB (3 nm) reference samples, using the identical experimental setups as the IrO$_2$ (111)/CoFeB. The signals can be fitted using only $V_{DL}^Y$ for $V_S$, and only $V_{FL}^Y$ for $V_A$. These results confirmed that the unconventional ST-FMR signals in Figure 2 of the main text originate uniquely from USHE of IrO$_2$ (111). The SOT efficiency of Ta was estimated as -0.05, which is consistent with previous reports.[50,51]

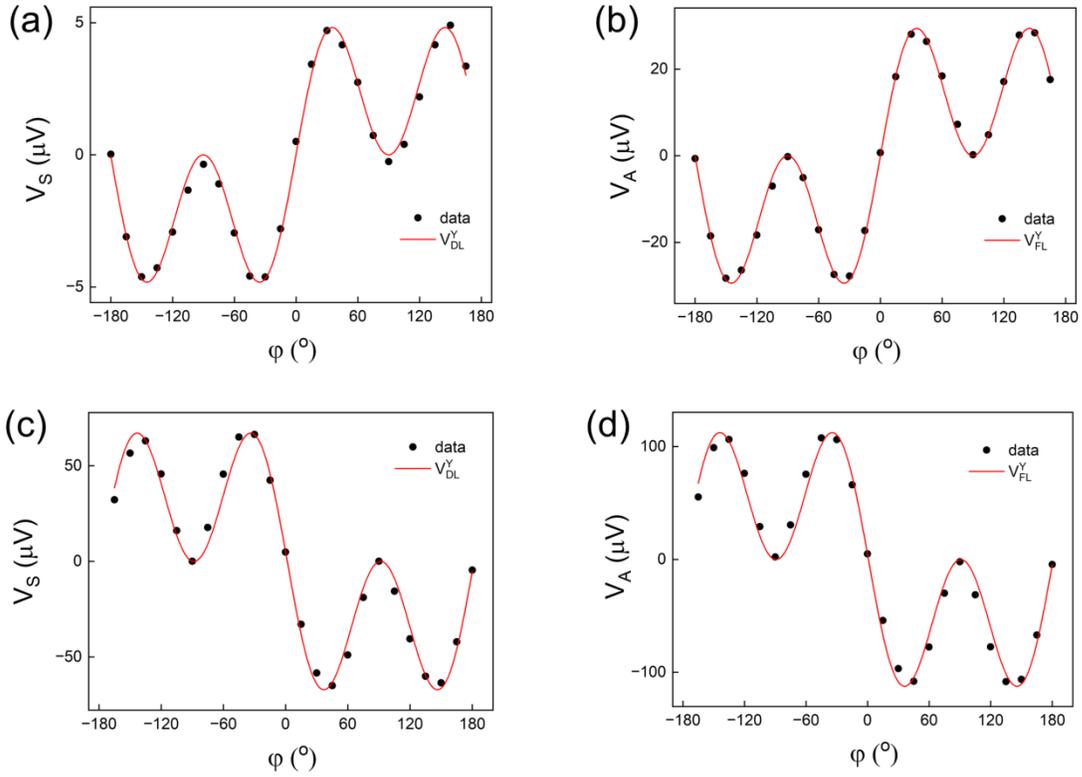

Figure S6. (a),(b) Symmetric and anti-symmetric ST-FMR signals of Ta/CoFeB heterostructure. (c),(d) Symmetric and anti-symmetric ST-FMR signals of $Bi_2Se_3$/CoFeB heterostructure.

## 4. Hall measurements of $IrO_2$ and $IrO_2$/CoFeB samples

Hall voltage of $IrO_2$ (111) film was measured using various AC current frequencies as well as DC current, with the results presented in Figure S7. It can be observed that the Hall voltage is independent of the current frequency.

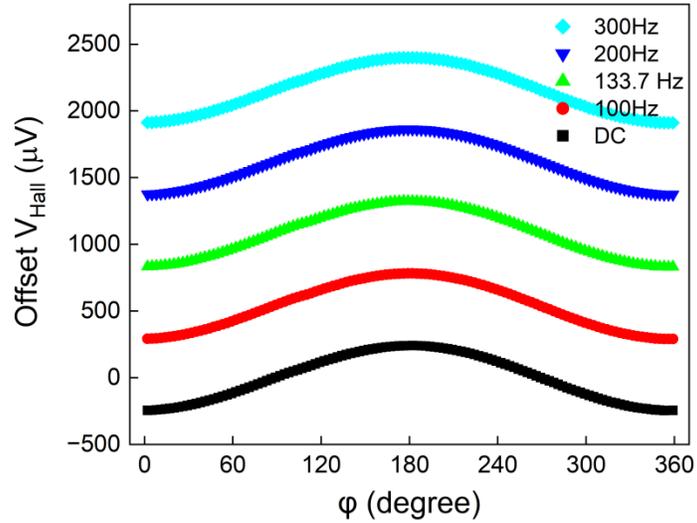

Figure S7. Angular dependence of Hall voltage of $IrO_2$ (111) (10 nm) at 5 T with current along $[\bar{1}\bar{1}2]$ direction at various frequencies. The curves are offset for better visibility.

Hall measurements were performed on an $IrO_2$ (111) (7 nm) sample as well as an $IrO_2$ (111) (10 nm)/CoFeB (5 nm) sample, as shown in Figure S8. APHE remains the dominant signal, exhibiting shifts with changes in current direction. These results are consistent with the observations for the $IrO_2$ (111) (10 nm) shown in Figure 3(d) in the main text.

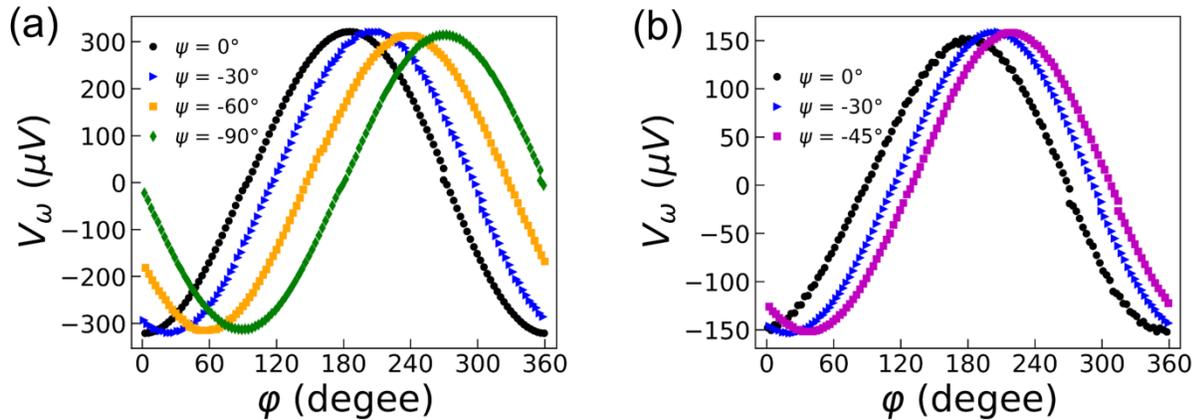

Figure S8. Hall voltage at 5 T for devices with different ψ angles of (a) $IrO_2$ (111) (7 nm) and (b) $IrO_2$ (111) (10 nm)/CoFeB (5 nm). ψ is defined in Figure 3(a) in the main text.

An example SPHE signal of $IrO_2$ (111) film measured at 50 K is shown in Figure S9. The signal can be separated to two regions by field magnitude. In the small field range (field < 4.5 T), the data follows a quadratic trend, whereas it shows a linear dependence on the large field range (field ≥ 4.5 T).

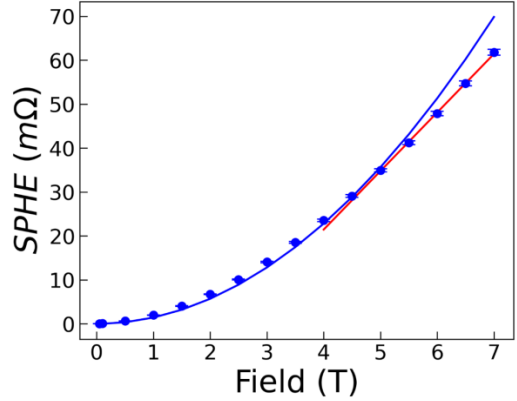

Figure S9. SPHE of $IrO_2$ (111) film. The result is fitted by a quadratic curve in the small field region and a linear curve in large field region.

There are three possible contributions to APHE: the intrinsic orbital effect, Berry curvature, and Lorentz force.[30] Assuming that the relaxation time is inversely proportional to temperature $T$,[52] these contributions are expected to scale with $T^2$, $T$, and $T^0$, respectively.[30] To investigate the contributions to the APHE signal of $IrO_2$ (111), the slopes of the APHE resistance ($R_{APHE}$) versus field ($B$) for different temperatures in Figure 4(b) of the main text were calculated, as shown in Figure S10. We used the function $R_{APHE}/B = c_2 T^2 + c_1 T + c_0$ to fit the data, where $c_2, c_1, c_0$ correspond to the intrinsic orbital effect, Berry curvature, and Lorentz force contributions, respectively. The fitted $c_2, c_1, c_0$ are -1.9×10$^{-5}$, 0.13, and 26.68, respectively, indicating the Lorentz force contribution is the dominant mechanism for the APHE.

Note that the fitting results may not precisely represent the three possible contributions. The significant differences among $c_2, c_1$ and $c_0$ primarily indicate that the temperature variation of APHE is small. The temperature variation may instead originate from the temperature-dependent Hall resistivity coefficients, as discussed in the report.[48]

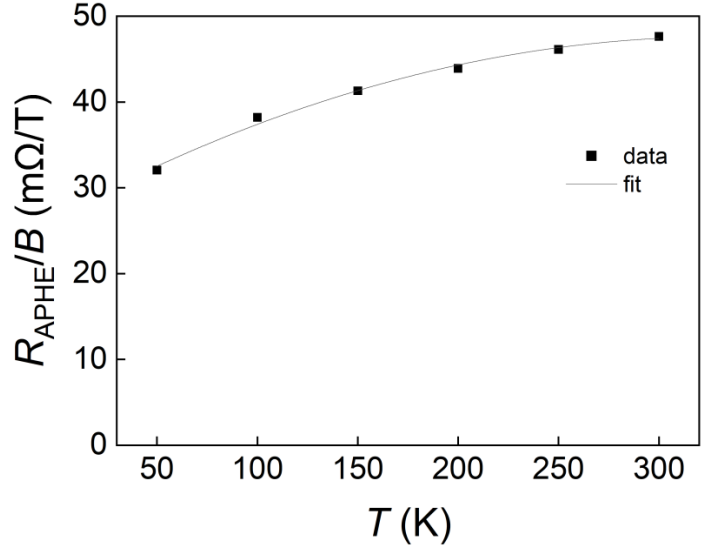

Figure S10. Temperature dependence of APHE magnitudes ($R_{APHE}/B$) of $IrO_2$ (111). The applied current is along $IrO_2$ $[\bar{1}\bar{1}2]$ direction.

Hall measurements were performed on $IrO_2$ (001)/CoFeB and $IrO_2$ (111)/CoFeB, which are depicted in Figure S11. These results confirm the APHE is originated from the crystal symmetry of $IrO_2$ (111).

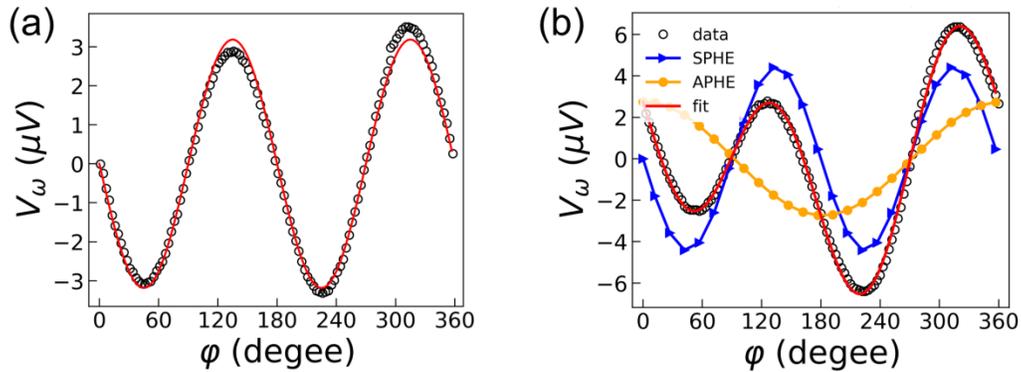

Figure S11. Hall voltage of $IrO_2$ (001)/CoFeB (a) and $IrO_2$ (111)/CoFeB (b). $IrO_2$ (001) shows an absence of APHE whereas $IrO_2$ (111) exhibits sizeable APHE voltage in addition to the PHE voltage from CoFeB and $IrO_2$ (111).

## 5. Symmetry analysis of USHE and APHE of IrO$_2$

The conductivity and spin Hall conductivity tensors are shown in Table SI. Due to the existence of $M_{(110)}, M_{(1-10)}, M_{(001)}$ and $C_{2z}$ symmetries, no USHE or APHE components are allowed for IrO$_2$ (001). In contrast, IrO$_2$ (111) only exhibits one mirror symmetry about the (1-10) plane ($M_{(1-10)}$). This low symmetry allows the existence of both USHE and APHE, as highlighted in Table SI.

Table SI. Conductivity and spin Hall conductivity tensors of IrO$_2$ (001) and IrO$_2$ (111) constrained by symmetry.

| | IrO$_2$ (001)<br>x = [100]<br>y = [010]<br>z = (001) | IrO$_2$ (111)<br>x = [-1-12]<br>y = [-110]<br>z = (111) |
|---|---|---|
| Symmetry | $M_{(110)}, M_{(1-10)}, M_{(001)}, C_{2z}$ | $M_{(1-10)}$ |
| $\sigma$ | $\begin{pmatrix} \sigma_{xx} & 0 & 0 \\ 0 & \sigma_{yy} & 0 \\ 0 & 0 & \sigma_{zz} \end{pmatrix}$ | $\begin{pmatrix} \sigma_{xx} & 0 & \sigma_{xz} \\ 0 & \sigma_{yy} & 0 \\ \sigma_{zx} & 0 & \sigma_{zz} \end{pmatrix}$ |
| $\sigma^x$ | $\begin{pmatrix} 0 & 0 & 0 \\ 0 & 0 & \sigma^x_{yz} \\ 0 & \sigma^x_{zy} & 0 \end{pmatrix}$ | $\begin{pmatrix} 0 & \sigma^x_{xy} & 0 \\ \sigma^x_{yx} & 0 & \sigma^x_{yz} \\ 0 & \sigma^x_{zy} & 0 \end{pmatrix}$ |
| $\sigma^y$ | $\begin{pmatrix} 0 & 0 & \sigma^y_{xz} \\ 0 & 0 & 0 \\ \sigma^y_{zx} & 0 & 0 \end{pmatrix}$ | $\begin{pmatrix} \sigma^y_{xx} & 0 & \sigma^y_{xz} \\ 0 & \sigma^y_{yy} & 0 \\ \sigma^y_{zx} & 0 & \sigma^y_{zz} \end{pmatrix}$ |
| $\sigma^z$ | $\begin{pmatrix} 0 & \sigma^z_{xy} & 0 \\ \sigma^z_{yx} & 0 & 0 \\ 0 & 0 & 0 \end{pmatrix}$ | $\begin{pmatrix} 0 & \sigma^z_{xy} & 0 \\ \sigma^z_{yx} & 0 & \sigma^z_{yz} \\ 0 & \sigma^z_{zy} & 0 \end{pmatrix}$ |

To clearly see the high symmetry of IrO$_2$ (001) and the low symmetry of IrO$_2$ (111), their lattice are depicted in Figure S12. IrO$_2$ (001) exhibits two mirror symmetries about two orthogonal planes ($M_{(110)}, M_{(1-10)}$), in addition to the mirror symmetry about the (001) plane and the ($M_{(001)}$) and a two-fold rotation symmetry about the [001] axis ($C_{2z}$). However, IrO$_2$ (111) only exhibits one mirror symmetry about the (1-10) plane ($M_{(1-10)}$).

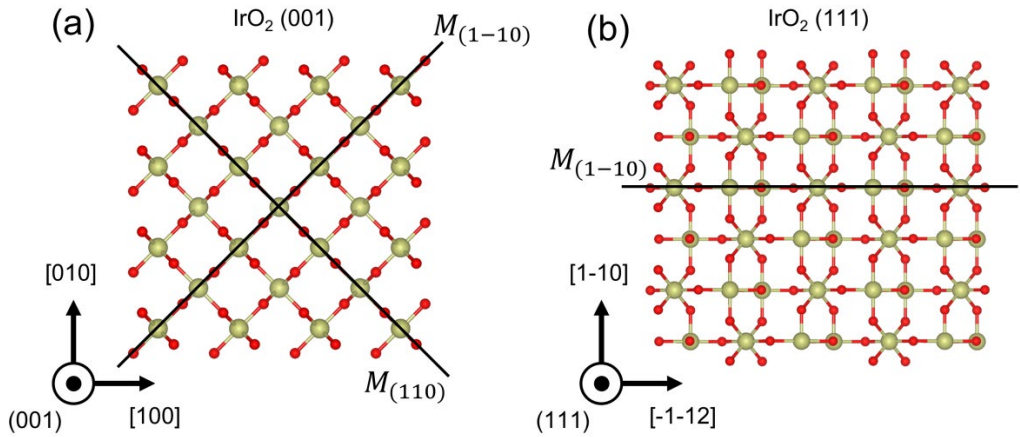

Figure S12. Lattice of IrO$_2$ (001) (a) and IrO$_2$ (111) (b).